\documentclass[
	twocolumn,
	amsmath,
	amssymb,
	prl,
	aps,
        superscriptaddress
	]{revtex4-2}
\usepackage{graphicx}
\usepackage{xcolor}
\usepackage{amsmath}
\usepackage{amssymb}
\usepackage{titlesec}
\usepackage{multirow}
\usepackage{lipsum}
\usepackage{comment}
\usepackage{babel}

\newcommand{\ds}[1]{\textcolor{black}{#1}}

\begin{document}

\title{Geometric control and memory in networks of hysteretic elements}

\author{Dor Shohat}
\affiliation{Department of Condensed Matter, School of Physics and Astronomy, Tel Aviv University, Tel Aviv 69978, Israel}
\affiliation{AMOLF, Science Park 104, 1098 XG Amsterdam, The Netherlands}
\author{Martin van Hecke}
\affiliation {AMOLF, Science Park 104, 1098 XG Amsterdam, The Netherlands}
\affiliation{Huygens-Kamerling Onnes Lab, Universiteit Leiden, PObox 9504, 2300 RA Leiden, The Netherlands}

\begin{abstract}
The response of driven frustrated media stems from  interacting hysteretic elements. 
We derive explicit mappings from 
networks of hysteretic springs to their abstract representation as interacting hysterons.
These maps reveal how the physical network controls the signs, magnitudes, symmetries, and pair-wise nature of the hysteron interactions. In addition,
strong geometric nonlinearities can produce pathways that require excess hysterons or even break hysteron models. Our results pave the way for metamaterials with geometrically controlled interactions, pathways, and functionalities, and highlight fundamental limitations of abstract hysterons in modeling {disordered} systems.

\end{abstract}

\maketitle

The response of driven disordered media, such as compressed crumpled sheets or sheared amorphous solids, forms pathways composed of sequential transitions between metastable states \cite{mungan2019networks,regev2021topology,keim2020global,bense2021complex,shohat2022memory}.
These pathways encode memories of past driving \cite{keim2019memory}, including its direction \cite{galloway2022relationships} and amplitude \cite{keim2011generic,paulsen2014multiple,mukherji2019strength,shohat2023dissipation}, and even computational capabilities \cite{bense2021complex,kwakernaak2023counting,liu2024controlled,treml2018origami}.
Understanding the connections between properties of a physical system and
its pathways is
crucial, both for the fundamental understanding of amorphous solids \cite{mungan2019networks,regev2021topology,kumar2022mapping,paulsen2024mechanical}, and for devising (meta)materials with targeted pathways, memory effects, or \textit{in-materia} computational capabilities \cite{liu2024controlled,sirote2024emergent,ding2022sequential,hyatt2023programming,el2024tunable}. 

\begin{figure}[t]
\centering
 \includegraphics[width=0.95\linewidth]{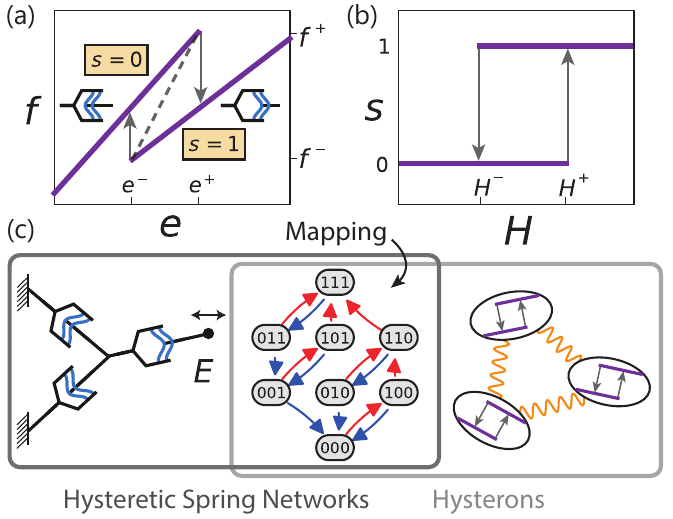}
\caption{(a) Force-displacement curve for a
hysteretic spring with two linear branches (purple);
pictograms represent
the configurations of a physical realization \cite{liu2024controlled}
(b) The binary phase $s:=0,1$ of a hysteron as a function of the external field $H$. (c) Relation between networks of hysteretic springs driven by displacement $E$ (left) and interacting hysteron models (right). Under certain conditions, networks and hysterons can be precisely mapped and produce the same transition-graph (center); however, in other cases, springs may follow non-hysteron pathways and vice versa. In the t-graph, states are represented with binary strings, and transitions with 
red (blue) arrows for increasing (decreasing) external driving.
} \label{fig:1}
\end{figure}

Physically, {these pathways are} often composed of {successive} flips of localized hysteretic elements such as beams, ridges, or particle clusters, which function as "material bits" \cite{shohat2022memory,keim2020global,bense2021complex,jules2022delicate,ding2022sequential}. Therefore, strictly binary hysteretic elements, known as hysterons \cite{preisach1935magnetische,van2021profusion}, are prime candidates
for modeling and designing memory effects and pathways \cite{mungan2019networks,lindeman2021multiple,keim2021multiperiodic,teunisse2024transition,lindeman2025generalizing}  (Fig.~\ref{fig:1}a-b). The strength of hysteron models is that they condense physical systems into a small set of parameters (switching thresholds) which characterize the material bits and their interactions \cite{van2021profusion,lindeman2021multiple}.
However, we lack a general link between these parameters and the underlying physics. Materializing targeted pathways from the hysteron model remains a challenge \cite{liu2024controlled,paulsen2024preprint}, while random hysteron parameters can produce non-physical responses or become ill-defined \cite{van2021profusion,baconnier2024proliferation,teunisse2024transition}.
Can we use hysteron models to describe and design physical systems composed of hysteretic elements?

To address these questions, we consider physical networks of hysteretic springs, and derive explicit mappings to the switching thresholds of the hysteron model (Fig.~\ref{fig:1}c).
For linear geometries, we map hysteretic spring networks to hysterons with pairwise interactions; for two dimensional networks, this mapping produces geometrically tunable, non-pairwise interactions. We leverage this geometric control to realize exotic pathways, including
multiperiodic orbits where the systems only returns to its initial state after multiple driving cycles \cite{keim2021multiperiodic,szulc2022cooperative}. 
However, when springs align or orient perpendicular to the drive, this mapping 
requires more hysterons than springs or may even break down:
exceptional
networks can exhibit pathways
beyond those of the hysteron framework.

Our work conveys two key insights. 
First, as exceptional alignment is unavoidable in large disordered systems, it may
lead to discrepancies with
hysteron models, and we identify key ingredients to improve mesoscopic models for amorphous media. Second, 
our work provides a precise, geometric, and general strategy to materialize targeted pathways in specific networks of
hysteretic elements, {whether electronic \cite{strukov2008missing,emmerich2024nanofluidic,yeo2025polytype}, pneumatic \cite{muhaxheri2024bifurcations,djellouli2024shell}, fluidic \cite{martinez2024fluidic} or mechanical \cite{sirote2024emergent,ding2022sequential,paulsen2024preprint}.}  
{This} paves 
the way for rationally designed (meta)materials with on-demand memories and computational capabilities \cite{hyatt2023programming,el2024tunable,paulsen2024mechanical,yang2023bifurcation}.

{\em Hysteretic springs, pathways and t-graphs.---}
We consider hysteretic elements, characterized by a
bilinear relation between two conjugate physical variables, such as current and voltage \cite{strukov2008missing,emmerich2024nanofluidic,yeo2025polytype}, pressure and volume or flow \cite{muhaxheri2024bifurcations,martinez2024fluidic,djellouli2024shell}, or, as we use here, force $f_i$ and displacement $e_i$ \cite{puglisi2000mechanics,liu2024controlled}:
\begin{equation}
f_{i}=(k^0_{i} +\Delta k_{i}s_{i}) e_{i}-g_{i}s_{i}~.
\label{eq:asymforce}
\end{equation}
Each hysteretic spring switches between 
its two phases
$s_i\!=\!0$ and $s_i\!=\!1$ when its displacement $e_i$ exceeds the local, {hysteretic} switching thresholds {$e_i^+>e_i^-$}; the stiffnesses of its branches are $k_i^0$ and $k_i^1 := k_i^0 + \Delta k_i$, and $g_i$ sets
the force jumps between branches (Fig.~\ref{fig:1}a) \cite{supplementary}.
\nocite{meeussen2023multistable,terzi2020state}

We connect $n$ of these springs in a network and follow their response to edge-applied driving {by controlling the overall deformation} $E$. We assume no simultaneous flips, and focus on slow driving with overdamped dynamics (Fig.~\ref{fig:1}c); other conditions, including spatially textured driving \cite{bense2021complex,sirote2024emergent}, race conditions \cite{van2021profusion}, and dynamic driving \cite{jules2023dynamical,lindeman2023competition}  introduce additional complexity. 
The response is piecewise smooth, but when
$e_i$ reaches its individual switching thresholds,
the $i^{th}$ element flips its phase $s_i: 0\! \leftrightarrow\!1$. These events cause a rapid change in the forces and extensions, which in turn may trigger additional flips (avalanches). The critical values of $E$ where transitions are triggered define the global, state dependent switching thresholds $E_i^\pm(S)$, where $S:=(s_1,s_2,\dots)$. The transitions and switching thresholds can be collected in the transition graph (t-graph),
which captures the response to any sequential driving protocol and encodes the memory capacity and capabilities (Fig.~\ref{fig:1}c) \cite{paulsen2019minimal,mungan2019networks,bense2021complex,van2021profusion}.

{\em Hysteron model.---} 
A simplified description of such networks - and complex media in general - is provided by hysterons, {abstracted hysteretic elements which are purely binary, and} driven by an external field $H$ \cite{van2021profusion,keim2021multiperiodic,lindeman2021multiple} (Fig.~\ref{fig:1}b).
A collection of interacting hysterons is defined by the switching thresholds $H_i^\pm(S)$ which specify when hysteron $i$ flips between phases $s_i=0,1$, and which determine the t-graph 
\cite{van2021profusion,teunisse2024transition,lindeman2021multiple}.
Often, the state dependent thresholds are modeled via pairwise interactions:
\begin{equation}
H_i^\pm(S) = h_i^\pm - \sum_j c_{ij}^\pm s_j, \mbox{  with  } c^\pm_{ii}=0~,
\label{eq:abstract_model}
\end{equation}
where $h_i^\pm$ are the {fixed} bare thresholds and $c_{ij}^\pm$ are interaction coefficients \cite{van2021profusion,bense2021complex,liu2024controlled}. Hysteron models lack a physical description for these parameters and
typically assign them randomly \cite{keim2021multiperiodic,szulc2022cooperative,teunisse2024transition,lindeman2025generalizing}; however, for specific parameters, race conditions and unphysical loops may occur \cite{baconnier2024proliferation}.

{\em Interactions and mapping.---}
A successful mapping between a network and the hysteron model requires that they exhibit the same pathways and the same switching thresholds.
In hysteretic spring networks, interactions are mediated by physical balance equations, e.g., force balance. 
The switching thresholds can then be determined as follows:
{\em (i)} freeze the state $S$ so that the force-displacement relations $f_i(e_i)$ are strictly linear;
{\em (ii)} use force balance to determine $e_i$ as function of the driving $E$;
{\em (iii)} determine state-dependent switching thresholds $E_i^\pm(S)$ by calculating the values of $E$ where $e_i\!=\!e_i^\pm$ (either analytically or numerically). 
Once $E_i^\pm(S)$ is determined, we can trivially
identify these with $H_i^\pm(S)$, and then determine whether the interactions are pairwise, top-down symmetric (i.e. $c_{ij}^+=c_{ij}^-$)), and their {precise} values. 
Hence, our goal is to map a given geometry of springs with their parameters
to the switching thresholds $E_i^\pm(S)$.

\begin{figure}[t]
\centering
\includegraphics[width=0.95\linewidth]{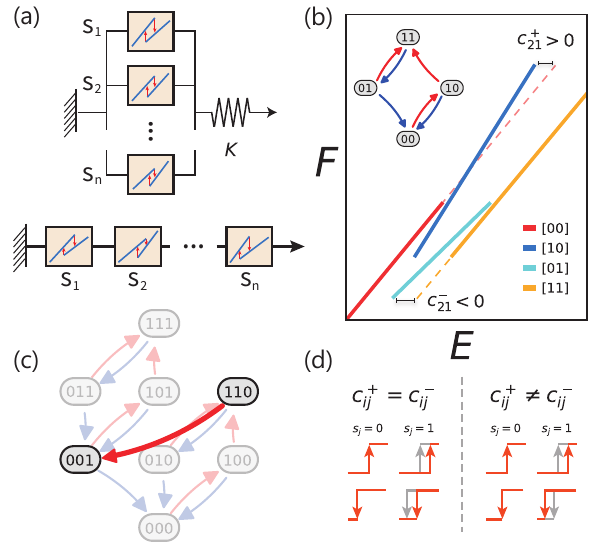}
\caption{(a) Hysteretic springs 
under controlled deformations interact
when coupled in parallel (spring $K$ mediates interactions, top) or in series (bottom). (b)
Collective response curves $F(E,S)$ for two serially coupled hysteretic springs, with endpoints representing state dependent switching thresholds.
The interaction coefficients follow from
comparing appropriate pairs of branches, e.g.,
$c_{21}^+=  E(f_2^+,(00))-E(f_2^+,(01))$ and
$c_{21}^-=  E(f_2^-,(01))-E(f_2^-,(11))$. Here these have opposite signs.
Inset: corresponding t-graph. (c) T-graph featuring a dissonant avalanche
(See \cite{supplementary} for details and parameters).
(d) Interactions that respect (left) and break (right) the up/down symmetry $c_{ij}^+=c_{ij}^-$ (arrows represent switching thresholds; orange: bare, grey: with interactions).
} \label{fig:2}
\end{figure}

{\em Linear networks.---} 
We first consider mechanical hysteretic springs
in serial or parallel
geometries: these can be mapped onto interacting hysterons
(Fig.~\ref{fig:2}a). The corresponding force balance equations are scalar and produce pairwise interactions (see Eq.~\ref{eq:abstract_model}). 
These can be worked out explicitly from the collective force-displacement curves $F(E,S)$ (Fig.~\ref{fig:2}b) \cite{supplementary}.
For serial coupling and controlled total displacement $E=\sum e_i$ \footnote{Serially coupled hysterons do not experience interactions under controlled force $F$ \cite{liu2024controlled}.}
we obtain
\begin{equation}\label{c_ser}
c_{ij}^\pm= -\frac{g_{j}}{n k_{j}^1} +
\frac{\Delta k_j}{n k_j^0  k_j^1}f_i^\pm~.
\end{equation}

For parallel coupling, an additional spring of stiffness $K$ mediates the interactions 
\footnote{For $K\rightarrow\infty$, parallel hysterons under controlled $E$ are non-interacting, and under controlled $F$ can be mapped to serially coupled hysterons under controlled $E$.}
and we obtain:
\begin{equation}\label{c_par}
c_{ij}^\pm=\frac{g_j-\Delta k_j e_i^\pm}{K}~.
\end{equation}

These expressions reveal
how geometry controls the interactions: serial coupling leads to negative interactions ($c_{ij}\!<\!0$) \cite{liu2024controlled}, while parallel coupling leads to positive interactions ($c_{ij}\!>\!0$).
For stiffness-symmetric springs (\(\Delta k = 0\)), interactions are up-down symmetric ($c_{ij}^+ = c_{ij}^-$) and global, i.e., each hysteron affects all others equally ($c_{ij} = c_{kj}$). In contrast, for \(\Delta k \neq 0\) this symmetry is broken - as seen in experiments \cite{bense2021complex} -  with large $\Delta k$ producing opposite signs of $c_{ij}^+$ and $c_{ij}^-$ (Fig.~\ref{fig:2}b,d).

These interactions control the pathways of the network. Interactions
may produce avalanches whose sequences of up ($u:0\rightarrow1$) and
down ($d:1\rightarrow0$) flips follow the signs of $c_{ij}^\pm$:
{\em(i)} In serial coupling of symmetric hysteretic springs ($\Delta k=0$), negative interactions produce avalanches that alternate between $u$ and $d$ and are of maximal length two \cite{liu2024controlled};
{\em(ii)} Stiffness-asymmetric springs can produce exotic, 'dissonant' avalanches like $duu$, where an decrease in {$E$ (or $H$)} leads to an increase in the magnetization $m:=\Sigma_i s_i$ or vice versa \footnote{Crucially, the dissonant avalanches here are sequential. If the first hysteron triggers more than one flip, we consider the avalanche ill-defined due to race conditions \cite{van2021profusion}.} (Fig.~\ref{fig:2}d) \cite{van2021profusion}; {\em(iii)} In parallel coupling, positive interactions drive monotonic avalanches (only $u$'s or $d$'s) of arbitrary length.
Recently, 
counter-snapping springs have been realized - where up (down) instabilities counter-intuitively lead to force jumps (drops) as opposed to
ordinary hysteretic springs  \cite{nicolaou2012mechanical, nicolaou2024metamaterials,ducarme_inprep}.
Such counter-snapping springs can be captured in our framework by controlling the signs of $g_j^{\pm}$, and can produce monotonic 
avalanches in serial geometries,
and  alternating avalanches in parallel geometries \cite{supplementary}.

Nevertheless, the range of
pathways in linear networks is limited. This is because all elements experience the same force (displacement) in serial (parallel) networks, so that their switching thresholds follow a fixed order set by the individual switching forces $f_i^{\pm}$ (displacements $e_i^{\pm}$) \cite{liu2024controlled}.
For example, if in a serial network $f_i^+<f_j^+$, 
then hysteron $i$ will always flip up before hysteron $j$. 

This limitation is reflected 
in the qualitative features of the t-graphs, which in particular cannot sustain a multiperiodic response.
We explore them numerically for three hysteretic elements in linear networks \cite{van2021profusion,liu2024controlled}, and find that out of hundreds of t-graphs, the vast majority obeys a hierarchical structure of nested loops known as loop-Return Point Memory (l-RPM) \cite{mungan2019structure}, which hinders the emergence of complex behavior \cite{supplementary}.
Hence, while the sign and magnitude of interactions can be controlled,
linear geometries severely limit the range of pathways and memory effects.

\begin{figure}[t]
\centering
\includegraphics[width=0.95\linewidth]{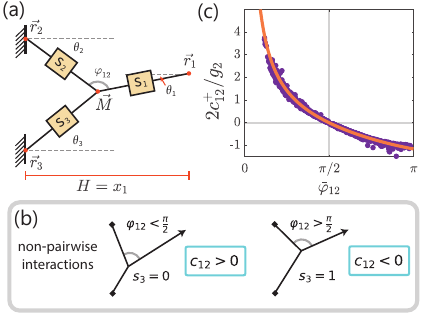}
\caption{(a) Trigonal hub with hysterons connected in 2D. $\vec{r}_1$ is externally pulled along the $\hat{x}$ axis.
(b) Strong non-pairwise
interactions, where flipping $s_3$ reverses the interactions between hysterons one and two.
(c) Comparison of the numerically obtained
$c^+_{12}$ (dots) and
approximate expression Eq.~(\ref{eq:C_ij_main}) (light curve) for
$10^4$ random samples; here we have assumed the geometrical relation $\theta_1=\pi/2-2\varphi_{12}/3$ \cite{supplementary}.
}
    \label{fig:3}
\end{figure}

{\em 2D networks.---} Our understanding of the linear geometries suggests that 2D networks of hysteretic {elements} 
may open a route to more advanced 
pathways and memory effects. We consider the paradigmatic case of a trigonal hub, where the two fixed and one moving corner of a triangle $\vec{r}_i=(x_i,y_i)$  connect to a central, freely moving point $\vec{M}$ via three symmetric hysteretic springs with $k_i=1$ (Fig.~\ref{fig:3}a). {We deform the hub by slowly driving $H=x_1$.}
Since the force balance on $\vec{M}$ is vectorial and angle dependent, the flipping order is no longer subordinate to the ordering of $f_i^{\pm}$ or $e_i^\pm$, allowing a far greater range of pathways, including multiperiodic responses.
We use this vectorial force balance to
derive exact expressions for the interaction coefficients, and find that
the ratio $g/e^{\pm}$ determines their form.
For small $g/e^{\pm}$, the interactions are pairwise and controlled by the angles between the springs, while for
large $g/e^{\pm}$, geometric non-linearities produce non-pairwise interactions and can even lead to a break down of the mapping from springs to hysterons.

When $g/e^{\pm}\ll1$,
the changes in angles {following an instability} are small and the interactions 
can be expressed geometrically:
\begin{align}
c_{ij}^{\pm}\approx z_i ~g_j\frac{\cos{\varphi}_{ij}}{\cos{\theta}_i}~,
\label{eq:C_ij_main}
\end{align}
where the geometrical factor $z_i$
arises from the connectivity of the trigonal hub ($\left[z_1,z_2,z_3\right]=\left[\frac{1}{2},1,1\right]$),
${\varphi}_{ij}$ is the angle between $s_i$ and $s_j$, and all angles are evaluated at the instability of hysteron $i$.
This expression has a clear geometrical interpretation.
First, the jump between the branches, $g_j$, provides the typical scale for the strength of interactions, as in the linear networks. Second, the factor $1/\cos\theta_i$ represents how global driving is coupled to stretching of hysteron $i$. Third, the factor $\cos\varphi_{ij}$ captures the geometric coupling: consistent with our findings for
serial and parallel couplings,
this factor approaches -1 (1) for ${\varphi}_{ij}=0$
(${\varphi}_{ij}=\pi$), and approaches zero for perpendicular hysterons.

When $g/e^{\pm}\not\ll1$, the angle changes during flipping events are significant. 
Interactions are neither pairwise nor up-down symmetric, and depend on the 
initial states. {For example, 
$c_{12}^{\pm}$ will depend on the phase of $s_3$} \cite{supplementary}. Yet, Eq.~\ref{eq:C_ij_main} still holds geometrical intuition: e.g., flipping $s_3$ may change ${\varphi}_{12}$. If as a result ${\varphi}_{12}$ crosses $\pi/2$, {$c_{12}^{+}(s_3=0)$ and $c_{12}^{+}(s_3=1)$} have opposite signs, presenting a strong deviation from pairwise interactions (Fig.~\ref{fig:3}b; see Suppl. Video 1).

We complement our analytical analysis with numerical simulations, where we determine $\vec{M}$ using overdamped dynamics: $\dot{\vec{M}}={\gamma^{-1}}\sum_i \vec{f}_i$, where $\gamma$ is a large damping constant and the forces are given by Eq.~(\ref{eq:asymforce}).  Starting from any state $S$ we quasi-statically drive $H$, follow $\vec{M}$ and identify instabilities whenever $e_i=e_i^{\pm}$. We find that the numerically calculated interaction coefficient $c_{12}^{+}$
closely matches the geometrical expression Eq.~(\ref{eq:C_ij_main}) (Fig. \ref{fig:3}c) \cite{supplementary}.

\begin{figure}[t]
\centering
\includegraphics[width=0.9\linewidth]{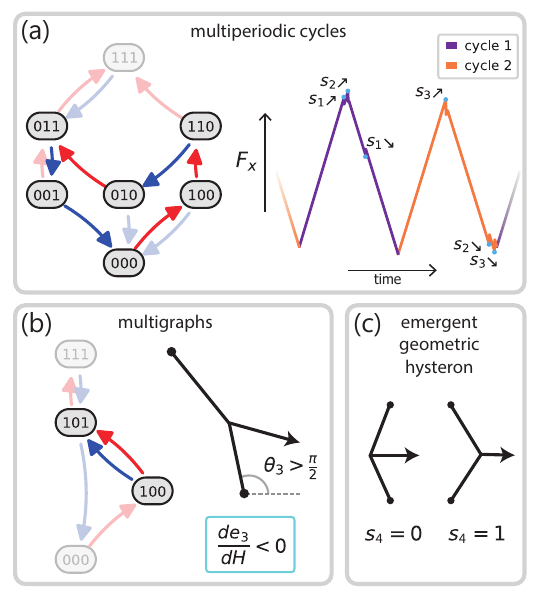}
\caption{(a) T-graph showing a multiperiodic cycle with periodicity $T=2$ (left);
Starting from state $(000)$ and driving $H$ cyclically, the system returns to $(000)$ after $T=2$ driving cycles \cite{supplementary}. Multiperiodicity is also apparent in the mechanical response (right), the force $F_x$ measured at $\vec{r}_1$ as a function of time. (b) Multigraph (left); Starting from $(100)$, either increasing or decreasing $H$ leads to the same transition $s_3=0\rightarrow1$; This happens when $e_3$ is perpendicular to the driving. Decreasing $H$ stretches $e_3$ if $\theta_3>\pi/2$ (right); (c) Illustration of an emergent, geometric $4^{th}$ hysteron. The state $S=(s_1,s_2,s_3)$ is fixed while the hub exhibits global, bistable buckling.
}
    \label{fig:4}
\end{figure}

{\em 2D t-graphs.---} The range of t-graphs that can be realized by 2D networks is huge, and
we use our numerical scheme to generalize earlier t-graph sampling protocols \cite{van2021profusion,liu2024controlled} to explore their diversity.
First, we find that the
trigonal hub can easily host multiperiodic cycles with a periodicity $T=2$ (Fig.~\ref{fig:4}; see Suppl. Video 2). Such cycles are a hallmark of complex behavior
in driven amorphous systems, break l-RPM, and cannot be realized in linear geometries {(because their
switching thresholds follow a fixed order).}

Second, the trigonal hub produces pathways where starting from a certain stable state $S$, either increasing or decreasing the drive $H$ triggers the \textit{same} transition. This violates the assumption of all hysteron models that
transitions under increased (decreased) driving are initiated
by a hysteretic {element}
flipping up (down). However, this effect can easily be understood geometrically, and is a general feature of higher dimensional networks: If the angle ($\theta_i$) between a 
spring 
and the driving $H$ is around $\pi/2$, both an increase and a decrease of $H$ can lead to stretching of this spring
thus triggering the same flipping (Eq.~\ref{eq:C_ij_main}). {The corresponding t-graph is a multigraph, where multiple directed transitions connect the same two states} (Fig.~\ref{fig:4}b; see Suppl, Video 3) \footnote{Multigraphs can occur
in abstract hysteron models via avalanches \cite{van2021profusion}, but
here neither of the multitransitions are avalanches.}.
Similarly, both for $H\rightarrow \infty$
and $H\rightarrow -\infty$, the
springs
get stretched and all switch to phase 1 --- hence, there is no guarantee that the springs will ever
reach $s_i=0$.

Third, 2D networks can give rise to additional hysteretic degrees of freedom associated with  buckling (Fig.~\ref{fig:4}c). This geometric degree of freedom can be mapped to an additional (fourth)
hysteron, for example by defining its phase via {the leaning direction} of the buckled structure \cite{lindeman2023competition}. Such geometric hysterons produce a wide variety of t-graphs \cite{supplementary}. 

For large random networks, we expect the numbers of springs perpendicular to the driving and of emergent hysterons to proliferate \cite{pettinari2024elasticity}. This suggests that for increasingly large systems, the mapping from spring networks to hysteron models progressively fails.

{\em Conclusion and Outlook.---}
Networks of hysteretic elements allow to investigate the links between real space configurations and abstract hysteron models. We have shown how geometry controls interactions, leading to a general, geometric strategy to materialize targeted pathways in networks of hysteretic elements. {At the same time, the exceptional 2D geometries clarify that spurious alignment between elements and between elements and driving, may break the description of physical systems by
hysteron models. In large networks, these effects could potentially be enhanced by non-affine deformations \cite{ellenbroek2009non}.} This breakdown implies that previous insights on the mechanics of amorphous solids, derived from hysteron models, {may require} reconsideration; 
points towards the importance of exceptionally aligned geometries for future descriptions of amorphous solids \cite{kumar2022mapping,lindeman2021multiple,nicolas2018deformation,puglisi2000mechanics}; and
suggests that physical systems exhibit
a wider range of responses and memory effects than shown in the hysteron model \cite{sirote2024emergent,jules2022delicate}.

While we focused {here} on controlled deformations, slow driving, and non-degenerate conditions, {we note} that stress control {(instead of strain) \cite{jules2022delicate}}, textured driving \cite{bense2021complex,sirote2024emergent},
dynamic driving \cite{jules2023dynamical,lindeman2023competition}, or race conditions \cite{van2021profusion}, can significantly extend the range of physically realizable pathways, {and further probe} the validity of hysteron models \cite{lindeman2023competition,teunisse2024transition}.

We briefly mention directions for future research. 
First, straightforward extensions include networks of 
mechanical elements with other {hysteretic} degrees of freedom
(e.g. shear or rotation) \cite{hyatt2023programming,kamp2024reprogrammable,paulsen2024preprint}. 
Second, the complex, non-pairwise interactions in 2D networks may provide a route to understanding the emergence of glassy dynamics in large networks \cite{shohat2023logarithmic,shohat2025emergent,cyclic_inprep}. 
Third, 
{physical hysteretic elements prohibit} the occurrence of spurious loops, where  after an instability, the system cannot find a stable state but instead gets trapped in a infinite cycle \cite{baconnier2024proliferation}. Such loops
overwhelm hysterons with arbitrary switching thresholds \cite{van2021profusion,keim2021multiperiodic,teunisse2024transition}, and also arise in coupled spin models \cite{eissfeller1994mean}.
Hence, physical models allow to define ensembles of hysterons that lead to well-defined dissipative behaviors.
Finally, our explicit expressions for the geometrically controlled interactions suggest that solving the inverse problem, i.e., translating a targeted pathway, t-graph, or set of switching thresholds to a specific network, is now within reach.

\section{Acknowledgements}
We thank Yoav Lahini, Paul Baconnier, Lishuai Jin, Colin Meulblok, and Margot Teunisse for insightful discussions. We are also grateful to Muhittin Mungan and Yair Shokef for a careful reading of the manuscript. This work was funded by the European Research Council Grant ERC-101019474. D.S. acknowledges support from the Clore Israel Foundation.

\bibliography{bibli.bib}

~\newpage
~\newpage

\onecolumngrid

\section{Supplemental Information}

\section{1. \ds{Bilinear hysteretic springs}}
\begin{figure}[b]
    \centering
    \includegraphics[width=0.75\textwidth]{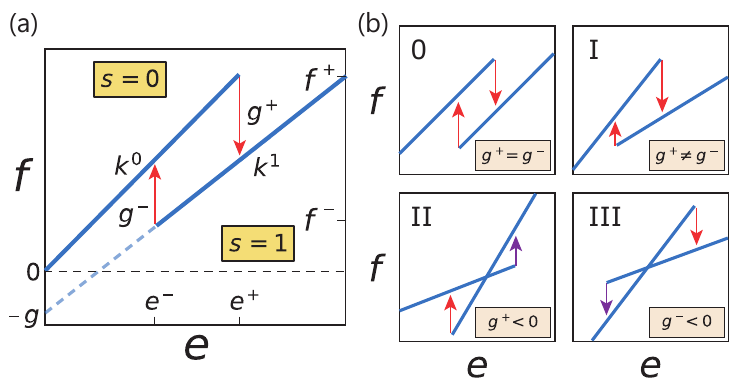}
    \caption{(a) Illustration of a hysteretic spring and its parameters. (b) Four types of \ds{hysteretic springs} classified by their force drops: \textbf{type 0} where the branches are symmetric and $g^+\!=\!g^-\!>\!0$; \textbf{type I} where the branches are asymmetric and $g^+\!\neq\! g^-\!>\!0$; \textbf{type II} where the branches cross each other and $g^+\!<\!0$; and \textbf{type III} where the branches cross each other and $g^-\!<\!0$.}
    \label{fig:sup_types}
\end{figure}

We consider mechanical \ds{hysteretic springs} where each of the two branches of the force-extension curve is linear (Fig. \ref{fig:sup_types}a). Their stiffness at $s_i\!=\!0$ is $k^0_i$, and at
$s_i\!=\!1$ is $k_i^1:=k_i^0\!+\!\Delta k_i$, so that
$\Delta k_i$ tunes the stiffness asymmetry between the $s_i=0$ and $s_i=1$
branches. These \ds{hysteretic springs} are assumed to have zero rest length (
the $s_i=0$ branch crosses through $(e_i,f_i)=(0,0)$), the $s_i=1$ branch
features an offset $g_i$. These \ds{springs} switch branch when the critical extension \ds{thresholds} $e_i^\pm$ are exceeded; specifically, when $e_i\uparrow e_i^+$, $s_i:0\rightarrow1$; similarly, $e_i\downarrow e_i^-$, $s_i:1\rightarrow0$.
These critical switching extensions define critical switching forces (the forces just before the instability):
$f_i^+=k_i^0e_i^+$ and $f_i^-=k_i^1e_i^--g_i$.
Finally, for $s_i\!:\!0\!\rightarrow\!1$ ($1\!\rightarrow\!0$), the force jump $g_i^+\!=\!g_i-\Delta k_i e_i^+$ ($g_i^-\!=\!g_i-\Delta k_i e_i^-$).

We can now classify \ds{hysteretic springs} based on their mechanical properties. We refer to symmetric \ds{elements} with $\Delta k=0$, or $g^+\!=\!g^-\!>\!0$, as {\textbf{type 0}}. Asymmetric \ds{elements} with $\Delta k\neq0$, and both $g_i^+\!>\!0$ and $g_i^-\!>\!0$, are referred to as {\textbf{type 1}}. Such \ds{hysteretic springs} with unequal natural stiffness can easily be realized \cite{meeussen2023multistable}.
Recently, bistable elements with self-crossing hysteresis loops have also been realized \cite{nicolaou2012mechanical,ducarme_inprep}, yielding \ds{hysteretic springs} where either
$g^+\!<\!0$ or $g^-\!<\!0$. We refer to these, respectively, as \textbf{type II} and \textbf{type III}. We require that the overall dissipation is positive ($\oint f_i \,de_i\geq0$). This classification is summarized in Fig. \ref{fig:sup_types}b.

We note that \ds{hysteretic springs} based on two other conjugated variables can similarly be classified, connected, and studied. These include electronic elements with hysteretic voltage-current responses \cite{strukov2008missing,emmerich2024nanofluidic,yeo2025polytype}; inflatable elements with hysteretic pressure-volume curves \cite{djellouli2024shell,muhaxheri2024bifurcations}; or fluidic devices with hysteretic pressure-flow curves \cite{martinez2024fluidic}.

\begin{figure}[t]
    \centering
    \includegraphics[width=0.75\textwidth]{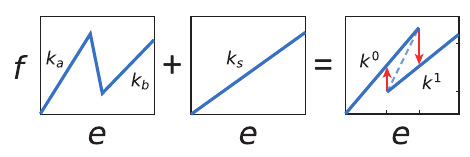}
    \caption{Creating a \ds{hysteretic spring} - a tri-linear element with a stiffness $k_a$ in the first branch and $k_b$ in the third branch, is coupled in series to a soft spring $k_s$. This result in a hysteretic response as described in the text.
    }
    \label{fig:sup_creating}
\end{figure}

Physically, \ds{hysteretic springs can be} constructed by combining an element with a non-monotonic force-displacement curve, as shown in Fig. \ref{fig:sup_creating}. We consider a trilinear element, coupled to a spring of stiffness $k_s$. Describing the trilinear element requires five parameters: The stiffness of the first branch $k_a>0$; the stiffness of the third branch $k_b>0$; the intersection of (the continuation of) the third branch with the $y$ axis $\tilde{g}$; the force at the first kink $f^+$; and the force at the second kink $f^-<f^+$.

\ds{When the trilinear element and the spring are} coupled in series, the first branch reads

\begin{align}
f^0=k^0 e,
\end{align}
and the third branch
\begin{align}
f^1=k^1 e-g,
\end{align}
where we conveniently defined $k^0\!\equiv\!\left(\frac{1}{k_a}+\frac{1}{k_s}\right)^{-1}$, $k^1\!\equiv\!\left(\frac{1}{k_b}+\frac{1}{k_s}\right)^{-1}$, and $g\equiv \frac{k^1\tilde{g}}{k_b}$.

We now consider the endpoints of these branches. This occurs when the force reaches $f^\pm$. We denote these critical extension thresholds $e^+\!\equiv\! e(f^0\!=\!f^+)$ and $e^-\!\equiv\! e(f^1\!=\!f^-)$. The condition for an emergent hysteresis loop is simply that
\begin{align}
\Delta e\equiv e^+ -e^-=\frac{f^+}{k^0}-\frac{f^-+g}{k^1}>0.
\end{align}

\section{2. \ds{Hysteron models:} transition graphs, scaffolds, avalanches, and scrambling}

\ds{The hysteron model considers $n$ purely binary, abstracted hysteretic elements, subjected to an external field $H$ \cite{van2021profusion}. Each element, a hysteron, has a binary phase $s_i=0,1$ and the state of the system is $S:=(s_1,s_2,\dots)$. Without interactions, hysteron $i$ switches up when $H$
is increased above the bare switching threshold $h_i^+$, and down when $H$ is decreased below the threshold $h_i^-\!<\!h_i^+$. Interactions are
encoded by introducing
state dependent switching thresholds $H_i^\pm(S)$,
namely that the switching thresholds of one hysteron depends on the phases of the others
\cite{teunisse2024transition,lindeman2021multiple}. Naturally, the complexity of pathways exhibited by a collection of hysterons depends on the number of free parameters $p$ in the model, in essence the number of independent switching thresholds it has. At most, this number can reach $p=n\cdot 2^{n}$  for $n$ hysterons (each state of their $2^n$ states requires $n$ independent critical thresholds). Often interactions are simplified and introduced via a linear pairwise dependence 
\cite{van2021profusion}:
\begin{equation}
H_i^\pm(S) = h_i^\pm - \sum_j c_{ij}^\pm s_j, \mbox{  with  } c^\pm_{ii}=0~,
\label{eq:abstract_model}
\end{equation}
where  $c_{ij}^\pm$ are interaction coefficients. This reduces the number of free parameters to $p=2n^2$; assuming 'up-down' symmetry ($c_{ij}^+\!=\!c_{ij}^-$), $p$ further reduces to $n^2+n$ (Fig.~2a); and assuming reciprocity ($c_{ij} \!=\! c_{ji}$) reduces $p$ even further to $(n^2\!+\!3n)/2$ \cite{keim2021multiperiodic,szulc2022cooperative}.
Finally, in the absence of interactions (the so-called Preisach model)
$H_i^\pm(S)\!=\!h_i^\pm$ and $p\!=\!2n$  \cite{terzi2020state}.
The choice of symmetries (and therefore the number of free parameters) varies between different studies.}

The t-graphs of coupled hysteron models feature nodes
that represent the stable states $S:=(s_1,s_2,\dots)$ and edges that
represent the transitions from state $S$ to state $S'$ that occur when $H$ is increased (decreased) past the relevant switching \ds{threshold}. We can think of
the t-graphs as composed of a scaffold that is dressed by (avalanche) transitions \cite{teunisse2024transition}.
To define the scaffold, we first associate each state with a pair of critical hysterons
$k^+(S)$ and $k^-(S)$, which are the
first hysterons that flip $0\rightarrow 1$ or $1\rightarrow 0$ when $H$ is increased or decreased:
\begin{equation}\label{eq:global_indices}
	\begin{split}
		k^+(S) &= \arg\min H_{i}^+(S) ~,\\
		k^-(S) &= \arg\max H_{i}^-(S)~,\\
	\end{split}
\end{equation}
where we note that the saturated states have only one critical hysteron.
The scaffold is the collection of critical hysterons $(k^\pm(S))$ at all stable states \cite{teunisse2024transition}. Finally, we denote the landing states obtained by flipping the respective critical hysterons as $S^+$ and $S^-$ .

We now distinguish two types of transitions between states: single step transitions and avalanches. Denote the value of $H$ that initiates the transition as $H^c$, then the stability of the landing states at $H^c$ determines the type of transition.
For single step transitions, the landing state $S^\pm$ is stable at $H^c$, and $S'$ is one of the landing states; for avalanches, the landing state $S^\pm$ is unstable at $H^c$, triggering additional transitions. However, note that all intermediate avalanches follow the scaffold. Hence. any possible transition is given by stitching together one or more steps specified by the scaffold \cite{teunisse2024transition}.

We distinguish two classes of scaffolds: Preisach scaffolds and scrambled scaffolds \cite{bense2021complex, van2021profusion}:
For Preisach scaffolds, the critical hysterons follow from a state-independent ordering
of the upper and lower switching \ds{thresholds}; hence if $k^+(000)=3$, then
$k^+(100)=3$. All possible Preisach scaffolds can easily be generated from the Preisach model of non-interacting hysterons \cite{terzi2020state}. The much more numerous scrambled scaffolds feature state dependent orderings of their
switching \ds{thresholds}, so that the critical hysterons of different states are decoupled, allowing, e.g.,
$k^+(000)=3$ and $k^+(100)=2$ \cite{van2021profusion}.
While t-graphs with avalanches based on a Preisach scaffold can exhibit non-trival properties, such
as breaking of return point memory \ds{or many transitory cycles that precede a periodic response} \cite{liu2024controlled}, many more advanced properties are believed to require
scrambling \cite{van2021profusion}.

\ds{Regarding the physical networks discussed in the main text, the scrambling property sets the difference between one-dimensional and two-dimensional systems. We showed that linear networks have a fixed ordering of the switching thresholds (set by $f_i^\pm$ or $e_i^\pm$). Thus, scrambling is prohibited, and the resulting t-graphs are limited in their complexity. In contrast, 2d networks allow for scrambling due to their vectorial balance equations, and accordingly host complex behaviors.}

\section{3. \ds{Mapping hysteretic springs in series to the hysteron model}}

We consider $n$ \ds{hysteretic springs} coupled in series. We denote their state $\boldsymbol{s}=\left(s_1,s_2,...,s_n\right)$, where $s_i=0,1$, representing the short and long configurations of the element respectively. We denote their individual forces and displacements $f_i$ and $e_i$. Serial coupling boundary conditions dictates

\begin{equation}
E=\sum_{i}e_i
\label{eq:sumdisp}
\end{equation}

\begin{equation}
f_{i}=f_{j}\quad \forall\ i,j
\label{eq:equalforce}
\end{equation}

We extend the existing framework of \cite{liu2024controlled} by considering a varying stiffness for each \ds{hysteretic spring}. The stiffness at the $s=0$ ($s=1$) branch is  $k_i^0$ ($k_i^1$). $\Delta k_i\equiv k_i^1-k_i^0$ is the stiffness difference that can take either positive or negative values. Conveniently, we write the force of the $i^{th}$ element as

\begin{equation}
f_{i}=(k_{i}^0+\Delta k_{i}s_{i})e_{i}-g_{i}s_{i}
\end{equation}

where the stiffness difference is coupled to $s_i$. $g_{i}$ is the intersection of the $s=1$ branch with the $y$ axis. The force drop $s=0\rightarrow1$ is $g_i^+=g_i-\Delta k_i e_i^+$, while the force drop $s=1\rightarrow0$ is $g_i^-=g_i-\Delta k_i e_i^-$. Next, force balance between elements gives

\begin{align}
e_{j}=\frac{f_i}{k_{j}^0+\Delta k_{j}s_{j}}+\frac{g_{j}s_{j}}{k_{j}^0+\Delta k_{j}s_{j}}
\end{align}

Plugging into Eq. \ref{eq:sumdisp} gives

\begin{align}
E=n\kappa^{-1}f_i+\sum_{j}\left(\frac{g_{j}s_{j}}{k_{j}^0 +\Delta k_{j}}-\Delta_j s_{j}f_i\right)
\label{eq:Uasym}
\end{align}

Where we have denoted $\kappa^{-1}=\langle\frac{1}{k}\rangle=\frac{1}{n}\sum_{j}\frac{1}{k_{j}}$, the average inverse stiffness at $S=0$; and $\Delta_j\equiv\frac{\Delta k_{j}}{k_{j}^0k_{j}^1}$, the stiffness asymmetry. We have absorbed all state dependencies to the sum. This sum term, which represents the interactions with other elements $j$, show that interactions do not exhibit 'up-down' symmetry. To this end, we can map the system to the hysteron model

\begin{align}
H_{i}^{\pm}=h_{i}^{\pm}-\sum_{j}c_{ij}^{\pm}s_{j}
\end{align}

Where $H$ is the external field; $h^{\pm}$ are the bare thresholds; $c_{ij}^{+}$ and $c_{ij}^{-}$, the interactions at $s_i=0$ and $s_i=1$ respectively (consistent with the notation for $e_{i}^{\pm}$).

We gauge out the self interaction (namely, we set $c_{ii}^{\pm}\equiv0$). This gauge is naturally consistent with the geometric interpretation for interactions presented in the main text. Altogether, the mapping gives

\begin{eqnarray}
H_{i}^{\pm}&=&\frac{E_{i}^{\pm}}{n},\\
h_{i}^{+}&=&\kappa^{-1}f_{i}^{+},\\
h_{i}^{-}&=&\kappa^{-1}f_i^-+\frac{1}{n}\left(\frac{g_{i}}{k_{i}^1}-\Delta_i f_i^-\right),\\
c_{ij}^{+}&=&-\frac{g_{j}}{n k_{j}^1} + \frac{\Delta_j}{n}f_i^+~\mbox{ for $i\ne j$},\\
c_{ij}^{-}&=&-\frac{g_{j}}{n k_{j}^1} + \frac{\Delta_j}{n}f_i^-~\mbox{ for $i\ne j$},\\
c_{ii}^{\pm}&=&0.
\end{eqnarray}

The difference between the up and down interaction coefficients $\Delta c_{ij}\equiv c_{ij}^{+}-c_{ij}^{-}$ reads simply

\begin{align}
\Delta c_{ij}= \frac{\Delta_j}{n}\left(f_i^+-f_i^-\right)
\label{eq:coup_dif_ser}
\end{align}

\section{4. \ds{Mapping hysteretic springs in parallel to the hysteron model}}

We consider $n$ \ds{hysteretic springs} coupled in parallel, and allow their stiffnesses $k_i$, $\Delta k_i$ to vary. If these are rigidly coupled so that $E=e_i$, there are no  interactions under controlled $E$. However, when this assembly of \ds{hysteretic springs} is coupled in series to a Hookean spring of stiffness $K$, length $\ell$, and rest length $\ell_0$, effective interactions arise. This geometry is described by:
\begin{eqnarray}
e_1&=&e_2=\cdots=e_n~, \label{eq:parequal}\\
E&=&e_i+\ell~,
\label{eq:pardisp}\\
K(\ell-\ell_0)&=&\sum_j f_j=\sum_j (k_{j}^0+\Delta k_{j}s_{j})e_{j}-g_{j}s_{j}~.
\label{eq:parforce}
\end{eqnarray}

From Eq.~\ref{eq:parforce} we isolate $\ell$ using
Eq.~\ref{eq:parequal}

\begin{align}
\ell=\ell_0+\sum_j \frac{k_j^0 e_i}{K} + \sum_j \frac{(\Delta k_j e_i-g_j)s_j}{K}
\end{align}

Plugging this expression to Eq.~\ref{eq:pardisp}, and using, we obtain:

\begin{align}
E-\ell_0=\left(\frac{\Sigma_j k_j^0 }{K}+1\right)e_i-\sum_j\frac{(g_j-\Delta k_j e_i) s_j}{K}~.
\end{align}

As before, we use the expression for the state dependent switching \ds{thresholds} to
obtain the parameters of the abstract hysteron model:

\begin{eqnarray}
H_{i}^{\pm}&=&E_{i}^{\pm}-\ell_0~,\\
h_{i}^{+}&=&\left(\frac{\Sigma_j k_j^0}{K}+1\right)e_i^+~,\\
h_{i}^{-}&=&\left(\frac{\Sigma_j k_j^1}{K}+1\right)e_i^--\frac{g_i}{K}~,\\
c_{ij}^{+}&=&\frac{g_j-\Delta k_j e_i^+}{K}~\mbox{ for $i\ne j$},\\
c_{ij}^{-}&=&\frac{g_j-\Delta k_j e_i^-}{K}~\mbox{ for $i\ne j$},\\
c_{ii}^{\pm}&=&0~.
\end{eqnarray}

Hence, ferromagnetic interactions whose strength is inversely proportional to the spring constant $K$ emerge between parallely coupled
\ds{hysteretic springs}. As for the serial case, including asymmetry ($\Delta k_j\neq0$) may introduce Anti-ferromagnetic interactions. Yet, similarly, these are by definition not capable of inducing AF avalanches. At each instability there's a force drop and consequentially all elements can only increase their length. 

The difference between the up and down interaction coefficients $\Delta c_{ij}\equiv c_{ij}^{+}-c_{ij}^{-}$ reads

\begin{align}
\Delta c_{ij}= -\frac{\Delta k_j}{K}\left(e_i^+-e_i^-\right)
\label{eq:coup_dif_par}
\end{align}

\section{5. Sampling t-graphs and avalanches}

We now discuss the properties and statistics of t-graphs and avalanches, sampled for $n=3$ and for different \ds{hysteretic spring} types and different linear geometries. We draw $>\!10^7$ random parameter sets $(e_i^{\pm}, k_i^0, k_i^1, g_i)$ for each type and each geometry (serial or parallel). Table \ref{tab1} summarizes our results.

The first block summarizes the coupling coefficients for each type and geometry. For asymmetric \ds{hysteretic springs}, the possible signs of $c_{ij}^{+}$ and $c_{ij}^{-}$ are nearly freely controllable. For type II (type III) they are limited as by construction $\Delta k>0$ ($\Delta k<0$), see Eqs. \ref{eq:coup_dif_ser} and \ref{eq:coup_dif_par}.

The second block shows the number of distinct t-graph topologies that we observed, and the number of these that obey l-RPM \cite{mungan2019structure}. \ds{Graphs that break l-RPM include avalanches \cite{liu2024controlled}, or transients of length $\tau=1,2$ \cite{bense2021complex} In the latter, the system requires $\tau$ transitory driving cycles before reaching a periodic response}. For type 0, it is possible to enumerate all possible t-graphs \cite{liu2024controlled}. For the other \ds{types}, we did not manage to exhaustively sample the high-dimensional parameter space, even with $>\!10^7$ parameter sets, so our numbers are lower bounds.

The third block summarizes the avalanche types. As discussed in the text, avalanches are inherited from the signs of interaction coefficients $c_{ij}^{\pm}$. We observe three types of avalanches. \textit{(i)} Alternating (alt) avalanches, which go back and forth between $u$ and $d$ flips. The Preisach scaffold that underlies all t-graphs in  linear
geometries limits these to be of length 2 (namely $ud$ or $du$) \cite{liu2024controlled}. They are driven by negative, antiferromagnetic interactions, namely $g\!>\!0$ in series or $g\!<\!0$ in parallel; \textit{(ii)} Dissonant (disso) avalanches, where increasing the driving $H$ leads to a decrease in the magnetization $m\!:=\!\sum{s_i}$, or vice versa ($udd$ or $duu$). These require a combination of negative and positive interactions, hence requiring 
asymmetry; \textit{(iii)} Monotonic (mono) avalanches which only involve $u$ transitions or $d$ transitions. These are driven by positive, ferromagnetic interactions, namely $g\!<\!0$ in series or $g\!>\!0$ in parallel. 

For example, type 0 \ds{hysteretic springs} in parallel have only positive interactions ($g\!>\!0$), and therefore only exhibit monotonic avalanches. As another example, for type II \ds{hysteretic springs} in series, we have $g^+\!<\!0$ and $g^-\!>\!0$. Therefore we find monotonic avalanches when increasing $H$, and alternating avalanches when decreasing $H$. Due to the asymmetry, dissonant avalanches can also occur when decreasing $H$. All the table entries can be worked out by considering the signs of interactions and force jumps and following this logic.

We note that longer mixed avalanches (which combine $u$ and $d$ flips) also exist yet are extremely rare. We observed a single length 4 $duud$ avalanche for type I \ds{hysteretic springs} in series (probability $\sim10^{-7}$). This avalanche occurs when  a dissonant avalanche is extended with one extra $d$ flip. By symmetry, $uddu$ avalanches should exist yet were not sampled. Furthermore, even longer avalanches are expected to exist in larger systems.

\begin{table*}[b]
\vspace*{-0.3cm}

\begin{tabular}{ |l||c|c||c|c||c|l|l|l|l|}
 \hline
 & \multicolumn{2}{|c||}{Coupling coefficients} & \multicolumn{2}{|c||}{Sampling statistics} & \multicolumn{5}{|c|}{Avalanches} \\
 \hline
 Network & sym.? & signs of $(c_{ij}^+ ~ c_{ij}^-)$ & \# t-graphs & \# l-RPM & type & $2\uparrow$ & $2 \downarrow$ & $3 \uparrow$ & $3\downarrow$ \\
 \hline
Serial 0 & Y & $(--)$ & 44 & 37 & alt &$ud$& $du$&.&.\\

Serial I & N & $(--,-+,+-,++)$ & $\geq$100 & $\geq$77 & alt / disso & $ud$ & $du$ & $udd$ & $duu$ \\

Serial II & N & $(--,+-,++)$ & $\geq$66 & $\geq$57 & mono $\uparrow$ / alt $\downarrow$ & $uu$ & $du$ & $uuu$ & $duu$ \\

Serial III & N & $(--,-+,++)$ & $\geq$66$^{\dagger}$ & $\geq$57$^{\dagger}$ & alt $\uparrow$ / mono $\downarrow$ & $ud$ & $dd$ & $udd$ & $ddd$ \\

\hline

Parallel 0 & Y & $(++)$ & 58 & 58 & mono & $uu$ & $dd$ & $uuu$ & $ddd$ \\

Parallel I & N & $(++,+-,-+,--)$ & $\geq$137 & $\geq$116 & mono & $uu$ & $dd$ & $uuu$ & $ddd$ \\

Parallel II & N & $(++,-+,--)$ & $\geq$52 & $\geq$45 & alt $\uparrow$ / mono $\downarrow$ & $ud$ & $dd$ & . $^\ddagger$ & $ddd$ \\

Parallel III & N & $(++,+-,--)$ & $\geq$52$^{\dagger}$ & $\geq$45$^{\dagger}$ & mono $\uparrow$ / alt $\downarrow$ & $uu$ & $du$ & $uuu$ & . $^\ddagger$ \\

 \hline
\end{tabular}
\caption{Sampling statistics as function of network geometry and \ds{hysteretic spring} type: symmetry of interaction coefficients; possible signs for $(c_{ij}^+ ~ c_{ij}^-)$; number of distinct t-graphs that we observed; number out of which that obeys l-RPM; types of avalanches observed (alternating, dissonant, or monotonic); avalanches of lengths 2 and 3, initiated by increasing ($\uparrow$) or decreasing ($\downarrow$) $H$. $^\dagger$ Types II and III are a mirror image of each other under top-down reflection ( $0\leftrightarrow1$ and $u\leftrightarrow d$ ), thus we sample only type II \ds{hysteretic springs} and assume that type III follow the same statistics; $^\ddagger$ No other avalanches were observed in our random sampling, yet we do not have analytic arguments that rule out their existence.
}\label{tab1}
\end{table*}

\section{6. Mapping the trigonal hub geometry}

\begin{figure}[t]
    \centering
    \includegraphics[width=0.4\textwidth]{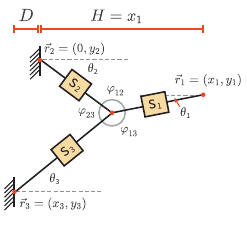}
    \caption{\ds{Geometry of the trigonal hub and the notation used throughout the derivation.}
    }
    \label{fig:sup_2d_geo}
\end{figure}

We consider three \ds{hysteretic springs} coupled in a 2D geometry as illustrated in Fig. \ref{fig:sup_2d_geo} in the main text. For simplicity we set $k_i=1$ and $\Delta k_i=0$. The \ds{springs} are freely rotating and can change their angles $\theta_i$ which are defined with respect to the horizontal axis. The external ends are pinned at $\vec{r}_i=(x_i,y_i)$, while the central point $\vec{M}$ is free to move on both axes. The horizontal displacement of the system $x_1$ is controlled externally. Without loss of generality, we set it as the external field $H\equiv x_1$. We also set $x_2=0$. The boundary conditions now depend on $\theta_i$. In the $\hat{x}$ axis we have

\begin{align}
H=e_1 \cos{\theta_1}+\frac{1}{2}(e_2 \cos{\theta_2}+e_3 \cos{\theta_3})-\frac{D}{2}
\label{eq:disp_x_2d}
\end{align}
where we denoted $D\equiv x_2-x_3$.

In the $\hat{y}$ axis we denote

\begin{align}
L_1=e_2 \sin{\theta_2}+e_3 \sin{\theta_3}
\label{eq:y_balance1}
\end{align}
\begin{align}
L_2=e_3 \sin{\theta_3} + e_1 \sin{\theta_1}
\label{eq:y_balance2}
\end{align}
where $L_1$ and $L_2$ are constants given by the geometry. This represents the fact that distances in the $\hat{y}$ axis are fixed (since we drive along $\hat{x}$).

Force balance on the central point gives

\begin{align}
f_1 \cos{\theta_1}=f_2 \cos{\theta_2}+f_3 \cos{\theta_3}\label{eq:force_x_2d}
\end{align}
\begin{align}
f_1 \sin{\theta_1}+f_2 \sin{\theta_2}=f_3 \sin{\theta_3}
\end{align}

\subsection{6.1. ${c_{12}^+}$ - a representative example}

We focus on the interaction $s_2$ exerts on the transition $s_1:0\rightarrow1$, and follow the full derivation.

First, we isolate $e_1$ from Eq. \ref{eq:force_x_2d}. Plugging into Eq. \ref{eq:disp_x_2d} we get

\begin{align}
H&=\frac{3}{2}e_1 \cos{\theta_1}-\frac{1}{2}g_1 s_1\cos\theta_1+\frac{1}{2}(g_2 s_2 \cos{\theta_2}+g_3 s_3 \cos{\theta_3})-\frac{D}{2}
\label{eq:H_u1}
\end{align}

Naively, one could suggest the following mapping to a hysteron model

\begin{align}
h_{1}^{+}=\frac{3}{2}e_{1}^{+}\cos{\theta_1}  \\
h_{1}^{-}=(\frac{3}{2}e_{1}^{+}-\frac{g_1}{2})\cos{\theta_1} \\
c_{12}=-\frac{g_{2}\cos{\theta_2}}{2}  \\
c_{13}=-\frac{g_{3}\cos{\theta_3}}{2}
\end{align}

However, this mapping turns to be incorrect. Its bare thresholds and interaction coefficients depend on the angles $\theta$. These are in turn determined by force balance in a non-linear manner, and obtain different values for different states $S$. As the angles are different in each state we expect $c_{ij}^+\neq c_{ij}^-$. Furthermore, activating $s_k=0\rightarrow1$ will change \textit{all} angles and therefore the coupling $c_{ij}^{\pm}$. Namely, interactions are no longer pairwise. To understand the geometry dependence of interactions, let us focus on the interaction that $s_2$ exerts on $s_1$ as a representative example.

The interaction coefficient $c_{12}^{+}$ includes changes in the bare threshold $h_{1}^{+}$ induced by flipping $s_2$. We denote the angles $\theta_i$ and $\theta_i'$ the angles at which $s_1$ reaches instability $e_1=e_1^+$, at $S=(00*)$ and $S=(01*)$ respectively. The interaction coefficient reads

\begin{align}
c_{12}^{+}=\underbrace{\frac{3}{2}e_1^+[\cos{\theta_1}-\cos{\theta_1'}]}_{\text{$\alpha_{12}^{+}$}}-\underbrace{\frac{g_{2}}{2}\cos{\theta_2'}}_{\text{$\beta_{12}^{+}$}}{\color{teal}+\underbrace{\frac{g_3s_3}{2}[\cos{\theta_3}-\cos{\theta_3'}]}_{\text{$\gamma_{12}^{+}$}}}
\label{eq:alpha_beta_decomp}
\end{align}

We shall show that, in fact, $c_{12}^{+}$ represents an intuitive and repeatable geometric dependence in the limit $g\ll e^{\pm}$. It is already apparent that in this limit $\gamma_{12}^{+}\ll\alpha_{12}^{+},\beta_{12}^{+}$. Thus for convenience we mark the smaller terms in teal.

To simplify the calculation we treat the three terms $\alpha$, $\beta$, $\gamma$ separately.

For the first term we have

\begin{align}
\alpha_{12}^{+}=\frac{3}{2}e_1^+[\cos{\theta_1}-\cos{\theta_1'}]=-3e_1^+\sin\bar{\theta}_1\sin(\frac{\theta_1-\theta_1'}{2})
\label{eq:ugly_alpha1}
\end{align}

where we used $\cos(a) - \cos(b) = -2 \sin\left(\frac{a + b}{2}\right) \sin\left(\frac{a - b}{2}\right)$. We denote $\bar{\theta}_1=\frac{\theta_1+\theta_1'}{2}$ the average angle.

To resolve $\alpha_{12}^{+}$, we now examine $e_1^+[\sin{\theta_1}-\sin{\theta_1'}]$. This is simply the vertical displacement of the central point, between the two instabilities. It's exact value can be obtained from force balance in the $\hat{y}$ axis. At the instability we have

\begin{align}
e_1^+ \sin{\theta_1}+e_2 \sin{\theta_2}=e_3 \sin{\theta_3}{\color{teal}-g_3s_3\sin{\theta_3}}\\
e_1^+ \sin{\theta_1'}+e_2' \sin{\theta_2'}-g_2\sin{\theta_2'}=e_3' \sin{\theta_3'}{\color{teal}-g_3s_3\sin{\theta_3'}}
\end{align}
for $s_2=0$ and $s_2=1$ respectively. By plugging Eqs. \ref{eq:y_balance1} and \ref{eq:y_balance2}, we get


\begin{align}
3e_1^+ \sin{\theta_1}=2L_2-L_1{\color{teal}-g_3s_3\sin{\theta_3}}\\
3e_1^+ \sin{\theta_1'}-g_2\sin{\theta_2'}=2L_2-L_1{\color{teal}-g_3s_3\sin{\theta_3'}}
\end{align}

Subtracting the two equations we get

\begin{align}
e_1^+[\sin{\theta_1}-\sin{\theta_1'}]=-\frac{g_2}{3}\sin{\theta_2'}{\color{teal}-\frac{g_3s_3}{3}[\sin{\theta_3}-\sin{\theta_3'}]}
\end{align}

Now, we explicitly subtract the sines on the LHS using $\sin(a) - \sin(b) = 2 \cos\left(\frac{a + b}{2}\right) \sin\left(\frac{a - b}{2}\right)$ to get

\begin{align}
e_1^+\sin(\frac{\theta_1-\theta_1'}{2})=-\frac{g_2}{6}\frac{\sin{\theta}_2'}{\cos\bar{\theta}_1}{\color{teal}-\frac{g_3s_3}{6}\frac{[\sin{\theta_3}-\sin{\theta_3'}]}{\cos\bar{\theta}_1}}
\label{eq:ugly_alpha2}
\end{align}

Altogether, plugging Eq. \ref{eq:ugly_alpha2} into Eq. \ref{eq:ugly_alpha1} gives

\begin{align}
\alpha_{12}^{+}=\frac{g_2}{2}\frac{\sin\bar{\theta}_1\sin{\theta}_2'}{\cos\bar{\theta}_1}{\color{teal}+\frac{g_3s_3}{2}\frac{\sin\bar{\theta}_1[\sin{\theta_3}-\sin{\theta_3'}]}{\cos\bar{\theta}_1}}
\label{eq:total_alpha}
\end{align}

We can now write the second term of Eq. \ref{eq:alpha_beta_decomp} as

\begin{align}
\beta_{12}^{+}=\frac{g_{2}}{2}\cos{\theta_2'}=\frac{g_{2}}{2}\frac{\cos\bar{\theta}_1\cos{\theta}_2'}{\cos\bar{\theta}_1}
\label{eq:total_beta}
\end{align}

Combining all contributions we get

\begin{align}
c_{12}^{+}&=\alpha_{12}^{+}-\beta_{12}^{+}+\gamma_{12}^{+}=-\frac{g_2}{2}\frac{\cos(\bar{\theta}_1+{\theta}_2')}{\cos\bar{\theta}_1}{\color{teal}+\frac{g_3s_3}{2}\left(\tan\bar{\theta}_1[\sin{\theta_3}-\sin{\theta_3'}]+[\cos{\theta_3}-\cos{\theta_3'}]\right)},
\end{align}
or if we subtract the sines explicitly
\begin{align}
c_{12}^{+}&=-\frac{g_2}{2}\frac{\cos(\bar{\theta}_1+{\theta}_2')}{\cos\bar{\theta}_1}{\color{teal}+g_3s_3\sin(\frac{\theta_3-\theta_3'}{2})[\tan\bar{\theta}_1\cos\bar{\theta}_3-\sin\bar{\theta}_3]}
\label{eq:C_no_approx}
\end{align}

We can now use the approximation $g\ll e^{\pm}$, namely that flipping \ds{a hysteretic spring} induces small length changes. In other words, this means that each instability also leads to small changes in the rest angles. Therefore, we can approximate $\theta_i\approx\theta_i'\approx\bar{\theta}_i$, or $\theta_i-\theta_i'\approx0$. This allows to simplify the above expression. First, we can drop the $g_3$ term. Second, we can replace $\bar{\theta}_1+\bar{\theta}_2=\pi-\bar{\varphi}_{12}$ with $\bar{\varphi}_{12}$, the angle formed between $s_1$ and $s_2$. This gives

\begin{equation}    
c_{12}^{+}\approx\frac{g_2}{2}\frac{\cos\bar{\varphi}_{12}}{\cos\bar{\theta}_1}
\label{eq:C_almost_final1}
\end{equation}

In this form, Eq. \ref{eq:C_almost_final1} can be understood through a simple geometrical intuition. First, the strength of the interaction is proportional to $\cos\varphi_{12}$. This angle determines how switching $s_2$ changes the length $e_1$. Next, the threshold $E_1^+$ diverges as $1/\cos\theta_1$. This represents how external driving stretches $e_1$. Finally, $g_2/2$ matches the analytical 1d solution when $\theta_1=0$ and $\varphi_{12}=\pi$ ($s_1$ is serially coupled to $s_{2,3}$ which are coupled in parallel).

We can even simplify further: for $\varphi_{12}\rightarrow 0$ mechanical balance give a simple relation $\theta_1=\pi/2-2\varphi_{12}/3$. It turns to hold for large angles on average, and we can approximate by an equation with one parameter only:

\begin{equation}
c_{12}^+\approx\frac{g_2}{2}\frac{\cos\bar{\varphi}_{12}}{\sin2\bar{\varphi}_{12}/3}
\label{eq:C_final}
\end{equation}

Thus, geometry dependence of interactions is systematic despite the breakdown of the simple mapping to the hysteron model.

This derivation also shows explicitly that interactions are inherently non-pairwise, and sheds light on their nature. When displacements at instabilities are large, namely $g\not\ll e^{\pm}$, the coefficient $c_{12}^+$ explicitly depends on $s_3$. Furthermore, flipping $s_3$ has another indirect effect - by varying the angle $\hat{\varphi}_{12}$. This provides intuition for the emergence of strong non-pairwise interactions. If flipping $s_3=0\rightarrow1$ changes $\hat{\varphi}_{12}$ such that it crosses $\pi/2$ when $s_3$ is flipped, the interaction $c_{12}^+$ can change its sign. In essence, changing the geometry can toggle between positive and negative (parallel-like and serial-like) coupling.

\subsection{6.2. All coefficients}

We can now derive all interaction coefficients in a similar manner. We begin with the up coefficients $c_{ij}^+$

Due to relabeling symmetry between $s_2$ and $s_3$, it suffices to consider $c_{21}^+$ and $c_{23}^+$.

Similarly to Eq. \ref{eq:H_u1}, we write $H$ as a function of $e_2$:

\begin{align}
H&=3e_2 \cos{\theta_2}-g_2 s_2\cos\theta_2+g_1 s_1 \cos{\theta_1}-g_3 s_3 \cos{\theta_3}+D
\label{eq:H_u2}
\end{align}

The first coefficient reads

\begin{align}
c_{21}^{+}=3e_2^+[\cos{\theta_2}-\cos{\theta_2'}]-g_{1}\cos{\theta_1'}{\color{teal}-g_3s_3[\cos{\theta_3}-\cos{\theta_3'}]}
\label{eq:alpha_beta_decomp}
\end{align}
where we have used the same notation of angles, this time with respect to the instability $s_2:0\rightarrow1$ and both states of $s_1$.
Neglecting the smaller term from $s_3$ gives our previous derivation, with relabeling between $s_1$ and $s_2$ and multiplying by 2:

\begin{align}
c_{21}^{+}\approx g_1\frac{\cos\bar{\varphi}_{12}}{\cos\bar{\theta}_2}
\label{eq:C_21_final}
\end{align}

The next coefficient reads

\begin{align}
c_{23}^{+}=3e_2^+[\cos{\theta_2}-\cos{\theta_2'}]+g_{3}\cos{\theta_3'}{\color{teal}+g_1s_1[\cos{\theta_1}-\cos{\theta_1'}]}
\end{align}

This expression is yet again similar to Eq. \ref{eq:alpha_beta_decomp}, but the second term has a plus sign instead of a minus sign, because $s_2$ and $s_3$ are at the opposite side of $s_1$, compared to the central point $\vec{M}$. Explicitly subtracting the cosines in the first term (as done above), now gives a minus sign for the same reason

\begin{align}
3e_2^+[\cos{\theta_2}-\cos{\theta_2'}]\approx-g_3\frac{\sin\bar{\theta}_2\sin{\theta}_3'}{\cos\bar{\theta}_2}
\end{align}

Altogether this gives

\begin{align}
c_{23}^{+}\approx g_3\frac{\cos\bar{\varphi}_{23}}{\cos\bar{\theta}_2}
\label{eq:C_23_final}
\end{align}
having plugged $\bar{\varphi}_{23}=\bar{\theta}_{2}+\bar{\theta}_{3}$.

This concludes the derivation for the up coefficients $c_{ij}^+$. For the down coefficients $c_{ij}^-$ we have exactly the same derivation, with the addition of a negligible term from $g_i$. For example, $c_{12}^-$ reads

\begin{align}
c_{12}^{-}&=\underbrace{\frac{3}{2}e_1^-[\cos{\theta_1}-\cos{\theta_1'}]}_{\text{$\alpha_{12}^{-}$}}-\underbrace{\frac{g_{2}}{2}\cos{\theta_2'}}_{\text{$\beta_{12}^{-}$}}{\color{teal}+\underbrace{\frac{g_3s_3}{2}[\cos{\theta_3}-\cos{\theta_3'}]}_{\text{$\gamma_{12}^{-}$}}}{\color{teal}-\underbrace{\frac{g_1}{2}[\cos{\theta_1}-\cos{\theta_1'}]}_{\text{$\delta_{12}^{-}$}}}
\end{align}

The last term will give a minute correction, which is negligible when we approximate $g\ll e^{\pm}$. Thus Eq. \ref{eq:C_almost_final1} holds here as well. Note that in this limit, the angles at both $u$ and $d$ instabilities are the same, therefore $c_{12}^{+}\approx c_{12}^{-}$.

Altogether we have a unified formula for all coefficients:

\begin{align}
c_{ij}^{\pm}\approx z_i g_j\frac{\cos\bar{\varphi}_{ij}}{\cos\bar{\theta}_i}~,
\label{eq:C_ij}
\end{align}
with $z_i$ being a geometrical factor that arises from the connectivity of the trigonal hub, $z_1=1/2$, $z_2=z_3=1$, and the angles assessed with respect to the instability $e_i=e_i^\pm$. As noted above, the interaction coefficients are top-down symmetric, i.e. $c_{ij}^+=c_{ij}^-$. \ds{Since at the small deformation limit $g\ll e^{\pm}$ the angles are essentially fixed, namely $\bar\theta_i=\theta_i=\theta_i'$ and $\bar\varphi_i=\varphi_i=\varphi_i'$, we have omitted this notation in the main text when referring to the angles evaluated at the instabilities.}

\section{7. Emergent geometric hysteron in 2D}

Aside of its three built-in \ds{hysteretic springs}, the trigonal hub geometry can host additional bistable degrees of freedom which arise from geometric non-linearities \cite{lindeman2023competition}. When $H\equiv x_1$ is manipulated externally, the edges of \ds{springs} $s_2$, $s_3$ can buckle to the left or to the right of the $\hat{y}$ axis so that two stable states with the central point $x_M<0$ and $x_M>0$ may emerge (see Fig. 4 in the main text). The transitions between these states are hysteretic \cite{lindeman2023competition}. Such buckling occurs if the rest lengths of the two \ds{hysteretic springs} $\ell_2$ and $\ell_2$ are together longer than the distance between them ($|\vec{r}_2-\vec{r}_3|$). So far, we have
considered type 0 \ds{hysteretic springs} with  rest length at $s_i=0$ ($s_i=1$) equal to $\ell^0_i=0$ ($\ell^1_i=g_i$); in the remainder of this section, we consider a positive rest-length $\ell_i^0$ for $s_i=0$, by taking the force displacement relation $f_i=e_i-\ell^0_i-g_i s_i$.

We now associate the global buckling mode {to} an additional $4^{th}$ hysteron, denoting $s_4=0$ ($s_4=1$) for states with $x_M<0$ ($x_M>0$). We include this hysteron in the numerical scheme for t-graph sampling, and identify changes in $s_4$ by tracking $x_M$.
We randomize the \ds{spring} parameters $\vec{r}_i$, $e_i^{\pm}$, $g_i$, and $\ell^0_i$, and compute the corresponding t-graphs systematically. Indeed, $s_4$ exhibits hysteretic transitions. Note that we do not limit self-intersection, namely the edges of the hub are free to cross each other without an additional steric interaction.

\begin{figure}
    \centering
    \includegraphics[width=0.4\textwidth]{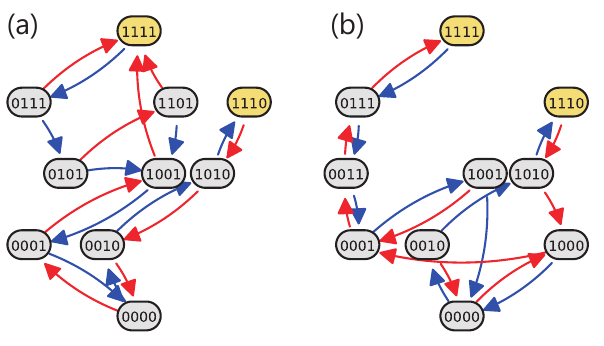}
    \caption{Two examples for transition graphs including the emergent $4^{th}$ hysteron (the notation follows $S={s_1,s_2,s_3,s_4}$. The saturated states, reached by extreme driving, are marked in yellow.}
    \label{fig:sup_4th}
\end{figure}

The range of transition graphs proliferates with the emergent degree of freedom $s_4$. Fig. \ref{fig:sup_4th} shows several examples. We stress, however, that an exhaustive sampling of the parameter space and graph properties is difficult, and a task we leave to future works. Instead, we highlight a peculiar feature that signals the breakdown of naive hysteron models. In the trigonal hub, saturated states (where all \ds{hysteretic springs} are at the 0 or 1 state) are not necessarily absorbing states which define a closed loop in the transition graph \cite{mungan2019structure}. Instead, the absorbing states (highlighted in Fig. \ref{fig:sup_4th}) are $(1110)$ and $(1111)$, namely when all three \ds{springs} are in the long state, and the whole trigonal hub is polarized to the left ($s_4=0$) or to the right ($s_4=1$).

\section{8. Figure parameters}

In this section we provide the parameters corresponding to the figures of the main text.

\textbf{Fig. 2b -} Two \ds{hysteretic springs} of type I are coupled in series. The corresponding parameters are: $e_i^+=[1,1.2]$; $e_i^-=[0.89, 0.58]$; $k_i^0=[1.5, 2.65]$; $k_i^1=[2.3, 1.57]$; $f_i^+=[1.52, 3.32]$; $f_i^-=[0.84, 0.3]$;  $g_i=[1.15 , 0.62]$. For panel c we calculate $c_{21}^\pm$ for different values of asymmetry $\Delta k_1$, while the rest of the parameters are fixed.

\textbf{Fig. 2dc -} Three \ds{hysteretic springs} of type I are coupled in series, and exhibit a dissonant avalanche. The corresponding parameters are: $e_i^+=[ 1.075,  1.02 ,  1.132]$; $e_i^-=[1.201,  1.357,  1.153]$; $k_i^0=[0.791,  1.465,  1.665]$; $k_i^1=[0.227,  1.667,  0.052]$; $f_i^+=[0.85 ,  1.494,  1.885]$; $f_i^-=[0.67 ,  0.42 ,  0.042]$;  $g_i=[-0.398,  1.843,  0.018]$. Note that here we allow $e_i^-\geq e_i^+$. As explained in previous sections, the hysteresis is emergent from the coupling of several elements, such that the hysteron model satisfies $h_i^-<h_i^+$.

\textbf{Fig. 3b -} Three \ds{hysteretic springs} of type 0 are connected in a trigonal hub. We randomize $10^4$ sets of positions and parameters. Importantly, we draw $e_1^+\in[2,6]$ and $g_i\in[0,0.4]$, thus maintaining the approximation used to derive the interaction coefficient $c_{12}^+$.

\textbf{Fig. 4a -} Three \ds{hysteretic springs} of type 0 are connected in a trigonal hub, and exhibit a multiperiodic cycle of period $T=2$, when driven along $H=[2.45,3.99]$. The corresponding parameters are: $x_i=[H,0,-0.339]$; $y_i=[-0.779,1.47,-1.442]$; $k_i=[1,1,1]$; $e_i^+=[2.722,2.087,1.943]$; $e_i^-=[2.426,1.984,1.719]$; $g_i=[0.066,0.379,0.358]$.

\textbf{Fig. 4b -} Three \ds{hysteretic springs} of type 0 are connected in a trigonal hub, and give rise to a multigraph, where from state $(100)$ driving either up or down leads to the same transition $(100)\rightarrow(101)$. The corresponding parameters are: $x_i=[H,0,1.05]$; $y_i=[-0.999,1.765,-1.669]$; $k_i=[1,1,1]$; $e_i^+=[1.53,2.471,1.457]$; $e_i^-=[1.344,2.386,1.425]$; $g_i=[0.457,0.031,0.15]$.

\section{9. Supplementary videos}

This section accompanies the supplementary movies, in which we visualize some exotic behaviors and pathways exhibited by the trigonal hub. The notation is as follows: \ds{hysteretic springs} are labeled in the same order as in the main text ($s_1$ on the right, $s_2$ on the top left, and $s_3$ on the bottom left); short \ds{springs} $s_i=0$ are colored in \textbf{blue}; long \ds{springs} $s_i=1$ are colored in \textbf{red}.

\textbf{Suppl. video 1 -} \ds{This video demonstrates strong non-pairwise interactions between hysteretic springs in the trigonal hub. Namely, the sign of $c_{12}^+$ depends on the state of $s_3$}. In the video, we examine the instability $s_1:0\rightarrow 1$ for both states of $s_2$. \ds{As explained in the text, comparing these two transitions gives the interaction coefficient. Here we focus only on the sign of the interaction coefficient. First, when $s_3=0$, \ds{element} $2$ promotes \ds{element} $1$ to flip earlier. We show side by side the instability for both states $s_2=0,1$. $s_1$ flips first on the right panel, therefore \ds{$c_{12}^+(s_3=0)>0$}. Next, we compare the same instabilities, but set $s_3=1$. In this case, \ds{element} $2$ hinders \ds{element} $1$ from flipping, and $s_1$ flips first on the left panel, therefore \ds{$c_{12}^+(s_3=1)<0$.} This signifies that the interaction has changed its sign.} Note that visually, in the former case $\varphi_{12}<\pi/2$, while in the latter case $\varphi_{12}>\pi/2$ as discussed in the main text. \ds{Here, the springs are symmetric (type 0), and} the corresponding parameters are: $x_i=[H,0.782,-0.266]$; $y_i=[2.754,1.702,-1.257]$; $k_i=[1,1,1]$; $e_1^+=2.621$; $g_i=[1.211,0.478,0.409]$.

\textbf{Suppl. video 2 -} This video visualizes the multiperiodic cycle \ds{shown in} Figure 4a in the main text \ds{for the trigonal hub}. \ds{Under periodic driving, the hub returns to its initial state after more than one driving cycle.} Starting from $(000)$, after one full cycle the hub arrives at $(010)$. It returns to the initial state $(000)$ only after a second driving cycle, thus realizing a multiperiodic cycle with periodicity $T=2$. The spring parameters appear in the previous section.

\textbf{Suppl. video 3 -} This video visualizes the multigraph shown in Figure 4b in the main text \ds{for the trigonal hub}. \ds{Here, starting from the same state, either increasing the extension of the hub leads to the same transition.} \ds{We start from $H=2.8$ and $(100)$. The video first shows the dynamics when $H$ increases, and then the dynamics when $H$ decreases. Both cases induce} the same transition $(100)\rightarrow(101)$. Notice that $s_3$ is rather perpendicular to the driving axis. When $H$ is decreased $\theta_3>\pi/2$ so that $e_3$ is stretched, in other words, $\frac{de_3}{dH}<0$ due to the geometry of the hub. The spring parameters appear in the previous section.

\end{document}